# Derivation of an Inverse Spatial Autoregressive Model for Estimating Moran's Index

Yanguang Chen

(Department of Geography, College of Urban and Environmental Sciences, Peking University, Beijing 100871, P.R. China. E-mail: chenyg@pku.edu.cn)

**Abstract**: Spatial autocorrelation measures such as Moran's index can be expressed as a pair of equations based on a standardized size variable and a globally normalized weight matrix. One is based on inner product, and the other is based on outer product of the size variable. The inner product equation is actually a spatial autocorrelation model. However, the theoretical basis of the inner product equation for Moran's index is not clear. This paper is devoted to revealing the antecedents and consequences of the inner product equation of Moran's index. The method is mathematical derivation and empirical analysis. The main results are as follows. First, the inner product equation is derived from a simple spatial autoregressive model, and thus the relation between Moran's index and spatial autoregressive coefficient is clarified. Second, the least squares regression is proved to be one of effective approaches for estimating spatial autoregressive coefficient. Third, the value ranges of the spatial autoregressive coefficient can be identified from three angles of view. A conclusion can be drawn that a spatial autocorrelation model is actually an inverse spatial autoregressive model, and Moran's index and spatial autoregressive models can be integrated into the same framework through inner product and outer product equations. This work may be helpful for understanding the connections and differences between spatial autocorrelation measurements and spatial autoregressive modeling.

# 1. Introduction

The spatial autocorrelation models based on Moran's index can be abstracted into a pair of equations. One equation is based on inner product of a size variable, and the other is based on the outer product of the same variable. The two equations can be employed to estimate Moran's index fast and generate canonical Moran's scatterplot (Chen, 2013). Where normalized Moran's scatterplot is concerned, the inner product equation generates scattered points, while the outer product equation generates a trend line. Based on the scattered points, another trend line can be formed. The slope of each trend line gives the value of Moran's index. The improved Moran's scatterplot possesses all the basic functions of the common Moran's scatter plot, the latter is invented by Anselin (1996). However, the normalized Moran scatter plot can show more geospatial information. The trend lines contain the information of both the spatial autocorrelation and spatial auto-regression. The distribution pattern of the scattered pointed along the trend lines indicates the significance level of spatial autocorrelation. The outer product equation can be derived from the formula of Moran's index based on a standardized variable and a globally normalized weight matrix. However, the inner product equation is not derivable from general principles or previous definition and requires further study before it will lead us to the underlying rationale of the pair of spatial autocorrelation equations. As a matter of fact, the inner product equation is obtained by considering one of the necessary conditions for defining Moran's index (Chen, 2013).

Spatial autocorrelation and spatial auto-regression represent two different sides of the same coin. The relationships between spatial autocorrelation and spatial auto-regression are easily understood. In the studies based on spatial statistics, spatial autocorrelation analysis tends to be mature (Cliff and Ord, 1969; Cliff and Ord, 1973; Geary, 1954; Getis, 2009; Griffith, 2003; Haining, 2009; Sokal and Oden, 1978; Sokal and Thomson, 1987; Tiefelsdorf, 2002), and spatial autoregressive modeling has a long history (Anselin,1988; Bennet, 1979; Cliff and Ord, 1981; Haining, 1979; Haining, 1980; LeSage, 1997; Pace and Barry, 1997; Upton and Fingleton, 1985; Ward and Gleditsch, 2008; Whittle, 1954). However, a lot of research does not mean the end of development of theory and methods in spatial analysis. The mathematical structure and statistical form of the relationship between spatial autocorrelation and auto-regression are not disclosed completely. In fact, based on the inner and outer product equations of Moran's index, the mathematical and statistical relationships between spatial autocorrelation and auto-regression can be examined from a novel prospective. The questions now are as follows. First, if the inner product equation of spatial autocorrelation cannot be derived from general principles, can we deduce it from the widely accepted empirical spatial autoregressive model? Second, how to reveal the mathematical structure and statistical nature of the relationships between the basic spatial auto-regression and the spatial autocorrelation based on Moran's index? This paper is devoted to deriving the inner equation of Moran's index from the simplest spatial autoregressive model. By doing so, we can obtain a series of useful or revealing results about spatial autocorrelation and spatial auto-regression. The rest parts are organized as follows. In Section 2, four sets of useful relations are derived from a simple spatial autoregressive model and the normalized Moran's index formula. In Section 3, a case analysis is made to lend an empirical support to the theoretical results. In Section 4, several related questions are discussed, and finally, in Section 5, the discussion will be concluded by summarizing the main points of this work.

## 2. Models

### 2.1 Basic formulae and relations

Spatial autocorrelation measurements and models can be simplified by means of standardized variable and normalized weight matrix. Suppose that there is a geographical region with $n$ elements such as cities. We want to measure the significance of spatial autocorrelation of these geographical elements. As we known, the basic spatial statistic is Moran's index. Based on standardized size variable and globally normalized spatial weight matrix, Moran's index can be express as a quadratic form (Chen, 2013)

$$I = \mathbf{z}^T \mathbf{W} \mathbf{z}, \tag{1}$$

where $I$ denotes Moran's index (Moran's $I$ for short), $\mathbf{z}$ is the standardized size vector by $z$-score, $\mathbf{W}$ is a $n \times n$ globally normalized symmetric spatial weight matrix, and the superscript T refers to transpose of matrix. The index was constructed by referring to Moran's idea about autocorrelation measure (Moran, 1950). According to the properties of globally normalized symmetric spatial weight matrix, the sum of element values in $\mathbf{W}$ is 1, and $\mathbf{W}^T = \mathbf{W}$. According to the properties of standardized variable, the mean of $\mathbf{z}$ is 0, and standard deviation of $\mathbf{z}$ is 1. Moreover, we have the following relation

$$\mathbf{z}^T \mathbf{z} = \mathbf{o}^T \mathbf{o} = n, \tag{2}$$

where $\mathbf{o} = [1, 1, \ldots, 1]^T$ represents a one vector consisting of $n$ ones. Equation (2) is the inner product of $\mathbf{z}$ and suggests that the correlation coefficient between a variable and itself is equal to 1.

Spatial autocorrelation models can be expressed as a pair of simple equations: one is outer product equation, and the other is inner product equation. Based on the outer product of **z**, we have the first characteristic function. Multiplying left equation (1) by **z** yields

$$\mathbf{z}\mathbf{z}^T\mathbf{W}\mathbf{z} = I\mathbf{z}. \tag{3}$$

This suggests that $I$ is the only nonzero eigenvalue of the matrix $\mathbf{z}\mathbf{z}^T\mathbf{W}$, and **z** is the corresponding eigenvector. It is easy to prove that Moran's index is the eigenvalue with the largest absolute value of $\mathbf{z}\mathbf{z}^T\mathbf{W}$. If $I>0$, it is the maximum eigenvalue of $\mathbf{z}\mathbf{z}^T\mathbf{W}$, and if $I<0$, it is the minimum eigenvalue of $\mathbf{z}\mathbf{z}^T\mathbf{W}$. Based on the inner product of **z**, we have the second characteristic function. An approximate characteristic function can be constructed as follows

$$\mathbf{z}^T\mathbf{z}\mathbf{W}\mathbf{z} = n\mathbf{W}\mathbf{z} = I\mathbf{z}, \tag{4}$$

which is a relation in theory. In fact, based on least squares method, there is a constant term in the above relation, thus equation (4) can be re-expressed as

$$\mathbf{z}^T\mathbf{z}\mathbf{W}\mathbf{z} = n\mathbf{W}\mathbf{z} = (\mathbf{o}^T\mathbf{W}\mathbf{z})\mathbf{o} + I\mathbf{z}, \tag{5}$$

in which $\mathbf{o}^T\mathbf{W}\mathbf{z} = (\mathbf{W}\mathbf{z})^T\mathbf{o}$ is a constant. The inner product $(\mathbf{W}\mathbf{z})^T\mathbf{o}$ equals the mean of $n\mathbf{W}\mathbf{z}$ (Appendix 1). This suggests that the mean of $\mathbf{W}\mathbf{z}$ is $(\mathbf{W}\mathbf{z})^T\mathbf{o}/n$. For observational data from the real world, equation (5) should be replaced by an empirical relation as below

$$\mathbf{z}^T\mathbf{z}\mathbf{W}\mathbf{z} = n\mathbf{W}\mathbf{z} = (\mathbf{o}^T\mathbf{W}\mathbf{z})\mathbf{o} + I\mathbf{z} + \mathbf{e}, \tag{6}$$

where **e** is a residuals set, representing an error term. It has been demonstrated that equation (3) and (4) can be used to generate a normalized Moran's scatterplot. Equation (4) generate scattered points, equation (3) generates a theoretical trend line, and equation (5) generates an empirical trend line. The slopes of the trend lines gives Moran's index value.

## 2.2 Derivations of spatial autoregressive coefficients

The spatial autocorrelation model can be derived from the spatial autoregressive model based on a standardized variable and a globally normalized spatial weight matrix. A simple spatial autoregressive model can be expressed as

$$\mathbf{z} = a\mathbf{o} + \rho\mathbf{W}\mathbf{z}, \tag{7}$$

where $\rho$ refers to auto-regressive coefficient. Equation (7) represent the pure theoretical expression for the simplest spatial autoregressive model. Note that, in the mathematical world, there is no error term. The corresponding empirical model can be rewritten as

$$\mathbf{z} = \hat{a}\mathbf{o} + \hat{\rho}\mathbf{W}\mathbf{z} + \boldsymbol{\varepsilon}, \tag{8}$$

where $\boldsymbol{\varepsilon}$ denotes residuals vector, the symbol "^" indicates estimated values of parameters. First of all, let's look at the reasoning process in the mathematical world. Equation (8) left multiplied by the transpose of vector $\mathbf{z}$ and $\mathbf{Wz}$, respectively, yields a pair of equations as follows

$$\mathbf{z}^T\mathbf{z} = a\mathbf{z}^T\mathbf{o} + \rho\mathbf{z}^T\mathbf{Wz}, \tag{9}$$

$$(\mathbf{Wz})^T\mathbf{z} = a(\mathbf{Wz})^T\mathbf{o} + \rho(\mathbf{Wz})^T\mathbf{Wz}. \tag{10}$$

Due to the mean of $\mathbf{z}$ is 0, the inner product of $\mathbf{z}$ and $\mathbf{o}$ is 0, that is $\mathbf{z}^T\mathbf{o} = \mathbf{o}^T\mathbf{z} = 0$. In terms of equations (1) and (2), equations (9) and (10) can be re-expressed as

$$0a + I\rho = n, \tag{11}$$

$$(\mathbf{Wz})^T\mathbf{o}a + (\mathbf{Wz})^T\mathbf{Wz}\rho = I. \tag{12}$$

According to Cramer's rule of linear algebra, a set of determinants based on the coefficient of equations (11) and (12) can be constructed as follows

$$A = \begin{vmatrix} n & I \\ I & (\mathbf{Wz})^T\mathbf{Wz} \end{vmatrix}, \quad B = \begin{vmatrix} 0 & n \\ (\mathbf{Wz})^T\mathbf{o} & I \end{vmatrix}, \quad C = \begin{vmatrix} 0 & I \\ (\mathbf{Wz})^T\mathbf{o} & (\mathbf{Wz})^T\mathbf{Wz} \end{vmatrix}.$$

Thus the formulae for the theoretical values of the regressive coefficients can be derived as below:

$$a = \frac{A}{C} = \frac{n(\mathbf{Wz})^T\mathbf{Wz} - I^2}{-I(\mathbf{Wz})^T\mathbf{o}}, \tag{13}$$

$$\rho = \frac{B}{C} = \frac{-n(\mathbf{Wz})^T\mathbf{o}}{-I(\mathbf{Wz})^T\mathbf{o}} = \frac{n}{I}. \tag{14}$$

Substituting equations (13) and (14) into equation (7) yields

$$\mathbf{z} = \frac{n(\mathbf{Wz})^T\mathbf{Wz} - I^2}{-I(\mathbf{Wz})^T\mathbf{o}} + \frac{n}{I}\mathbf{Wz}. \tag{15}$$

As indicated above, the mean of $\mathbf{z}$ is 0, and the mean of $\mathbf{Wz}$ is $(\mathbf{Wz})^T\mathbf{o}/n$. In light of the principle of linear regression analysis (Appendix 1), we have

$$a = \langle \mathbf{z} \rangle - \rho \langle \mathbf{Wz} \rangle = -\frac{1}{I}(\mathbf{Wz})^T\mathbf{o}, \tag{16}$$

where ⟨·⟩ means taking average value of $\mathbf{z}$ or $\mathbf{Wz}$. Comparing equation (13) with equation (16) yields

$$n(\mathbf{Wz})^T\mathbf{Wz} - I^2 = ((\mathbf{Wz})^T\mathbf{o})^2, \tag{17}$$

which is a pure theoretical relation. This relation will be used in the first process of deriving the formula of the least squares calculation. In fact, using transpose of $\mathbf{Wz}$ to multiply equation (5) left yields equation (17). Substituting equation (17) into equation (15) yields

$$\mathbf{z} = -\frac{1}{I}(\mathbf{Wz})^{\mathrm{T}}\mathbf{o} + \frac{n}{I}\mathbf{Wz}, \tag{18}$$

which we can converted into equation (5). This suggests that equation (5) can be derived from equations (18).

The mathematical world is a pure theoretical world, which is different from the real world and the computational world. In the mathematical world, the errors can be ignored. However, in the empirical world, residuals should be taken into account. Multiplying the transpose of vector $\mathbf{z}$ and $\mathbf{Wz}$ left by equation (8), respectively, yields another pair of equations as follows

$$\mathbf{z}^{\mathrm{T}}\mathbf{z} = a\mathbf{z}^{\mathrm{T}}\mathbf{o} + \hat{\rho}\mathbf{z}^{\mathrm{T}}\mathbf{Wz} + \mathbf{z}^{\mathrm{T}}\boldsymbol{\varepsilon}, \tag{19}$$

$$(\mathbf{Wz})^{\mathrm{T}}\mathbf{z} = a(\mathbf{Wz})^{\mathrm{T}}\mathbf{o} + \hat{\rho}(\mathbf{Wz})^{\mathrm{T}}\mathbf{Wz} + (\mathbf{Wz})^{\mathrm{T}}\boldsymbol{\varepsilon}. \tag{20}$$

For a regression model, the dependent variable is related to the residual term, but the independent variable is orthogonal to the residual term, that is

$$\mathbf{z}^{\mathrm{T}}\boldsymbol{\varepsilon} = \delta, \tag{21}$$

$$(\mathbf{Wz})^{\mathrm{T}}\boldsymbol{\varepsilon} = 0. \tag{22}$$

Here $\delta$ is a number to be calculated. From equations (19) and (20) it follows

$$0a + I\hat{\rho} = n - \delta, \tag{23}$$

$$(\mathbf{Wz})^{\mathrm{T}}\mathbf{o}a + (\mathbf{Wz})^{\mathrm{T}}\mathbf{Wz}\hat{\rho} = I. \tag{24}$$

According to Cramer's rule, three determinants based on the coefficient of equations (23) and (24) can be constructed as below:

$$A = \begin{vmatrix} n-\delta & I \\ I & (\mathbf{Wz})^{\mathrm{T}}\mathbf{Wz} \end{vmatrix}, \quad B = \begin{vmatrix} 0 & n-\delta \\ (\mathbf{Wz})^{\mathrm{T}}\mathbf{o} & I \end{vmatrix}, \quad C = \begin{vmatrix} 0 & I \\ (\mathbf{Wz})^{\mathrm{T}}\mathbf{o} & (\mathbf{Wz})^{\mathrm{T}}\mathbf{Wz} \end{vmatrix}.$$

Thus the formulae of model coefficients can be solved as

$$\hat{a} = \frac{A}{C} = \frac{(n-\delta)(\mathbf{Wz})^{\mathrm{T}}\mathbf{Wz} - I^2}{-I(\mathbf{Wz})^{\mathrm{T}}\mathbf{o}}, \tag{25}$$

$$\hat{\rho} = \frac{B}{C} = \frac{-(n-\delta)(\mathbf{Wz})^{\mathrm{T}}\mathbf{o}}{-I(\mathbf{Wz})^{\mathrm{T}}\mathbf{o}} = \frac{n-\delta}{I}. \tag{26}$$

According to the principle of linear regression analysis (Appendix 2), the autoregressive coefficient in equation (14) should be replaced by

$$\hat{\rho} = \frac{nR^2}{I}. \tag{27}$$

Thus, similar to equation (16), the constant term can be expressed as

$$\hat{a} = \langle \mathbf{z} \rangle - \hat{\rho} \langle \mathbf{Wz} \rangle = -\frac{R^2}{I}(\mathbf{Wz})^T \mathbf{o}. \tag{28}$$

Therefore, equation (18) should be replaced by

$$\mathbf{z} = -\frac{R^2}{I}(\mathbf{Wz})^T \mathbf{o} + \frac{R^2 n}{I} \mathbf{Wz} + \varepsilon. \tag{29}$$

Comparing equation (26) with equation (27) yields

$$\delta = n - nR^2 = n(1 - R^2). \tag{30}$$

In terms of equations (25), (28), and (30), the constant term can be re-expressed as

$$\hat{a} = \frac{R^2 n (\mathbf{Wz})^T \mathbf{Wz} - I^2}{-I(\mathbf{Wz})^T \mathbf{o}} = -\frac{R^2}{I}(\mathbf{Wz})^T \mathbf{o}. \tag{31}$$

Therefore, equation (29) can be rewritten as

$$\mathbf{z} = \frac{R^2 n (\mathbf{Wz})^T \mathbf{Wz} - I^2}{-I(\mathbf{Wz})^T \mathbf{o}} + \frac{R^2 n}{I} \mathbf{Wz} + \varepsilon. \tag{32}$$

From equation (31) we can derive the following relation

$$\frac{n(\mathbf{Wz})^T \mathbf{Wz} - I^2 / R^2}{(\mathbf{Wz})^T \mathbf{o}} = (\mathbf{Wz})^T \mathbf{o}. \tag{33}$$

Rearranging equation (33) yields

$$n(\mathbf{Wz})^T \mathbf{Wz} = ((\mathbf{Wz})^T \mathbf{o})^2 + \frac{I^2}{R^2}, \tag{34}$$

which will be used in the second process of deriving the formula of the least squares calculation.

## 2.3 Least squares algorithm for autoregressive coefficients

It can be proved that the least square method is suitable for estimating the values of autoregressive coefficients. Based on equation (7), the autoregressive coefficients defined in the mathematical world can be calculated by

$$\boldsymbol{\rho} = \left[ \begin{bmatrix} \mathbf{o}^T \\ (\mathbf{Wz})^T \end{bmatrix} \begin{bmatrix} \mathbf{o} & \mathbf{Wz} \end{bmatrix} \right]^{-1} \begin{bmatrix} \mathbf{o}^T \\ (\mathbf{Wz})^T \end{bmatrix} \mathbf{z} = \begin{bmatrix} \mathbf{o}^T \mathbf{o} & \mathbf{o}^T \mathbf{Wz} \\ (\mathbf{Wz})^T \mathbf{o} & (\mathbf{Wz})^T \mathbf{Wz} \end{bmatrix}^{-1} \begin{bmatrix} \mathbf{o}^T \mathbf{z} \\ (\mathbf{Wz})^T \mathbf{z} \end{bmatrix}, \tag{35}$$

where **ρ** refers to the coefficient vector. As indicated above, $\mathbf{o}^T\mathbf{z}=0$, $\mathbf{o}^T\mathbf{o}=n$, $(\mathbf{Wz})^T\mathbf{z}=I$. In terms of the principle of linear algebra, finding the inverse matrix of the square matrix in equation (35) yields

$$\boldsymbol{\rho} = \frac{1}{n(\mathbf{Wz})^T\mathbf{Wz} - ((\mathbf{Wz})^T\mathbf{o})^2} \begin{bmatrix} (\mathbf{Wz})^T\mathbf{Wz} & -(\mathbf{Wz})^T\mathbf{o} \\ -\mathbf{o}^T\mathbf{Wz} & \mathbf{o}^T\mathbf{o} \end{bmatrix} \begin{bmatrix} 0 \\ I \end{bmatrix}. \tag{36}$$

According to equation (17), equation (36) changes to

$$\boldsymbol{\rho} = \frac{1}{I^2} \begin{bmatrix} -I(\mathbf{Wz})^T\mathbf{o} \\ nI \end{bmatrix} = \begin{bmatrix} -\frac{1}{I}(\mathbf{Wz})^T\mathbf{o} \\ \frac{n}{I} \end{bmatrix}. \tag{37}$$

Apparently, the vector **ρ** gives two theoretical expressions: one is the constant term, and the other is the spatial autoregressive coefficients. Equation (37) corresponds to equation (13), equations (14), and equation (16).

For the empirical model, the vector expression of autoregressive coefficients is some different. By similar approach, the autoregressive coefficients defined in the computational world can be derived as

$$\hat{\boldsymbol{\rho}} = \frac{R^2}{I^2} \begin{bmatrix} -I(\mathbf{Wz})^T\mathbf{o} \\ nI \end{bmatrix} = \begin{bmatrix} -\frac{R^2}{I}(\mathbf{Wz})^T\mathbf{o} \\ \frac{R^2 n}{I} \end{bmatrix}. \tag{38}$$

The vector $\hat{\boldsymbol{\rho}}$ gives two empirical expressions: one is the constant term, and the other is the spatial autoregressive coefficients. Equation (38) corresponds to equation (27), equations (28), and equation (30). The derivation process and results suggests that the least squares regression can be utilized as a classic framework to estimate the values of spatial autoregressive model parameters.

## 2.4 Parameter bounds

Moran's index is a statistic measure and has its boundary values. Correspondingly, the spatial autoregressive coefficient has the boundary values related to the bounds of Moran's index. The value ranges of Moran's index and spatial autoregressive coefficient can be derived from the theoretical and empirical prospective. These theoretical results are based on the mathematical world. In terms of equations (3) and (14), we have

$$\frac{I}{n} = \frac{1}{\rho} = \frac{\mathbf{z}^T \mathbf{W} \mathbf{z}}{\mathbf{z}^T \mathbf{z}}. \tag{39}$$

According to the principle of Rayleigh quotient (Xu, 2021), we have the first value range of parameters as below

$$\lambda_{min} \leq \frac{I}{n} = \frac{1}{\rho} \leq \lambda_{max}, \tag{40}$$

where $\lambda$ is the eigenvalues of $\mathbf{W}$, $\lambda_{max}$ and $\lambda_{min}$ represent the maximum and minimum eigenvalues of $\mathbf{W}$, respectively. As a matter of fact, more than one method lead us to the conclusion that the value of Moran's index is confined by the maximum and minimum eigenvalues of spatial weight matrix (de Jong et al, 1984; Tiefelsdorf and Boots, 1995). In terms of equations (14) and (17), we have

$$\frac{((\mathbf{W}\mathbf{z})^T \mathbf{o})^2 + I^2}{n^2} = (\frac{1}{n}(\mathbf{W}\mathbf{z})^T \mathbf{o})^2 + \frac{1}{\rho^2} = \frac{\mathbf{z}^T \mathbf{W}^T \mathbf{W} \mathbf{z}}{\mathbf{z}^T \mathbf{z}}. \tag{41}$$

So we have the second value range of parameters as follows

$$\lambda^*_{min} \leq (\frac{1}{n}(\mathbf{W}\mathbf{z})^T \mathbf{o})^2 + \frac{I^2}{n^2} = (\frac{1}{n}(\mathbf{W}\mathbf{z})^T \mathbf{o})^2 + \frac{1}{\rho^2} \leq \lambda^*_{max}, \tag{42}$$

where $\lambda^*$ is the eigenvalues of $\mathbf{W}^T\mathbf{W}$, which is the inner product of $\mathbf{W}$, $\lambda^*_{max}$ and $\lambda^*_{min}$ represent the maximum and minimum eigenvalues of $\mathbf{W}^T\mathbf{W}$, respectively. Further, multiplying the transpose of vector $\mathbf{W}\mathbf{z}$ left by equation (3) yields

$$(\mathbf{W}\mathbf{z})^T \mathbf{z} \mathbf{z}^T \mathbf{W} \mathbf{z} = I(\mathbf{W}\mathbf{z})^T \mathbf{z} = I^2. \tag{43}$$

Multiplying the outer product of vector $\mathbf{z}$ by equation (5) left yields

$$n\mathbf{z}\mathbf{z}^T \mathbf{W} \mathbf{z} = \mathbf{z}\mathbf{z}^T \mathbf{o}(\mathbf{W}\mathbf{z})^T \mathbf{o} + I\mathbf{z}\mathbf{z}^T\mathbf{z} + \mathbf{z}\mathbf{z}^T \mathbf{e} = nI\mathbf{z}, \tag{44}$$

which is equivalent to equation (3). This suggests that, starting from equation (4) or equation (5), we can derive equation (43). In terms of equations (14) and (43), we have

$$\frac{I^2}{n} = \frac{n}{\rho^2} = \frac{\mathbf{z}^T \mathbf{W}^T \mathbf{z} \mathbf{z}^T \mathbf{W} \mathbf{z}}{\mathbf{z}^T \mathbf{z}}. \tag{45}$$

Then we have the third value range of parameters as follows

$$\lambda^{**}_{min} \leq \frac{I^2}{n} = \frac{n}{\rho^2} \leq \lambda^{**}_{max}, \tag{46}$$

where $\lambda^{**}$ is the eigenvalues of $\mathbf{W}^T\mathbf{z}\mathbf{z}^T\mathbf{W}$, which is the outer product of $\mathbf{W}\mathbf{z}$, $\lambda^{**}_{max}$ and $\lambda^{**}_{min}$ represent the maximum and minimum eigenvalues of $\mathbf{W}^T\mathbf{z}\mathbf{z}^T\mathbf{W}$, respectively. It can be proved that $\lambda^{**}_{min} = 0$, $\lambda^{**}_{max} = (\mathbf{Wz})^T\mathbf{Wz}$. This suggests that $I^2 \leq n(\mathbf{Wz})^T\mathbf{Wz}$, and $\rho^2 \geq n/(\mathbf{Wz})^T\mathbf{Wz}$.

The empirical value ranges of Moran's index and spatial autoregressive coefficient should be derived from the practical prospective. These results are based on the real world and computational world. In terms of equations (3) and (27), we have

$$\frac{I}{n} = \frac{R^2}{\hat{\rho}} = \frac{\mathbf{z}^T\mathbf{Wz}}{\mathbf{z}^T\mathbf{z}}. \tag{47}$$

Thus we have the first value range of parameters as below

$$\lambda_{min} \leq \frac{I}{n} = \frac{R^2}{\hat{\rho}} \leq \lambda_{max}, \tag{48}$$

where $\lambda$ is the eigenvalue of $\mathbf{W}$, $\lambda_{max}$ and $\lambda_{min}$ represent the maximum and minimum eigenvalues of $\mathbf{W}$, respectively. From equation (34) it follows

$$\frac{((\mathbf{Wz})^T\mathbf{o})^2 + I^2/R^2}{n^2} = (\frac{1}{n}(\mathbf{Wz})^T\mathbf{o})^2 + \frac{R^2}{\hat{\rho}^2} = \frac{\mathbf{z}^T\mathbf{W}^T\mathbf{Wz}}{\mathbf{z}^T\mathbf{z}}. \tag{49}$$

In terms of equations (27) and (49), we have the second value range of parameters as follows

$$\lambda^*_{min} \leq (\frac{1}{n}(\mathbf{Wz})^T\mathbf{o})^2 + \frac{I^2}{R^2 n^2} = (\frac{1}{n}(\mathbf{Wz})^T\mathbf{o})^2 + \frac{R^2}{\hat{\rho}^2} \leq \lambda^*_{max}, \tag{50}$$

where $\lambda^*$ is the eigenvalues of $\mathbf{W}^T\mathbf{W}$, $\lambda^*_{max}$ and $\lambda^*_{min}$ represent the maximum and minimum eigenvalues of $\mathbf{W}^T\mathbf{W}$, respectively.

## 3. Empirical analysis

### 3.1 Study area and data

This is a theoretical study on the relationships between spatial autocorrelation analysis and spatial autoregressive modeling. The results of mathematical derivation need to be verified by empirical analysis. After all, the success of sciences rests with their great emphasis on the role played by the interplay between quantifiable data and mathematical models (Louf and Barthelemy, 2014). To evaluate the theoretical reasoning results, three observational datasets will be employed to testify the models and relations given in Section 2. The study area is the well-known Beijing-Tianjin-Hebei (BTH) region of China. It is also termed Jing-Jin-Ji (JJJ) region in literature. The study area includes 35 cities in total ($n=35$). Concretely speaking, the BTH urban system consists of 13 prefecture level and above cities and 22 county-level cities. There are three sources of observational data. The spatial distances are measured by traffic mileage, which were extracted by ArcGIS. City sizes were measured by urban population based on the census in 2000 (the fifth census) and 2010 (the sixth census) as well as urban nighttime light (NTL) intensity. The data of NTL intensity is reflected by total number of NTLs within urbanized area of cities in BTH region (Chen and Long, 2021; Long and Chen, 2019). The spatial proximity is defined by the inverse distance function, $v_{ij}=1/r_{ij}$, where $r_{ij}$ denotes the traffic mileage between cities $i$ and $j$. Consequently, the spatial contiguity matrix can be expressed as $\mathbf{V}=[v_{ij}]=[1/r_{ij}]$, in which the diagonal elements are defined as zero. Globally normalizing $\mathbf{V}$ yields a spatial weight matrix $\mathbf{W}$, and the summation of the elements in $\mathbf{W}$ equals 1.

## 3.2 Results and analysis

The empirical analysis can be implemented from the following three aspects. First, testify the inherent relationships between the spatial autocorrelation model and corresponding spatial autoregressive model. Second, examine the constant term (intercept) and autoregressive coefficient (slope). Third, verify the key formulae and typical relations derived in last section. The analytical process is illuminated step by step as follows.

First of all, the parameter values can be estimated for spatial autocorrelation model and spatial autoregressive model by using the least squares calculations. The size measures, as indicated above, are the logarithms of city population and nighttime light intensity. The analysis process comprises five steps. **Step 1**: taking logarithms of city size variables in given time. Thus, we have $z=\ln(x)$, in which $x$ denotes the original variable, and $z$ refers to the logarithmic variable. **Step 2**: normalizing the spatial contiguity matrix. Thus we have $\mathbf{W}=\mathbf{V}/V_0$, where $\mathbf{V}$ is spatial contiguity matrix, $V_0$ is the sum of the elements in the $\mathbf{V}$, and $\mathbf{W}$ refers to the globally normalized spatial weight matrix. **Step 3**: estimating Moran's index and the related parameter values. Making regressive analysis based on equation (6) yields the estimated values of the parameters of spatial autocorrelation model. The independent variable is $\mathbf{z}$, and the dependent variable is $n\mathbf{Wz}$. So, the intercept is $(\mathbf{Wz})^T\mathbf{o}$, and the slope is the estimated value of Moran's index, $\hat{I}$. This step has been illustrated in a companion paper. **Step 4**: estimating the spatial autoregressive coefficient and the related parameter values. Making regressive analysis based on equation (8) yields the estimated values of the parameters of spatial autoregressive model. In this case, the independent variable is $\mathbf{Wz}$, and the dependent variable is $\mathbf{z}$ (Figure 1). So, the intercept is $\hat{a}$, and the slope is $\hat{\rho}$. **Step 5**: verifying the basic parameter relations. For example, for NTL intensity in 2010, Moran's index is about $I=0.1248$, the corresponding spatial autoregressive coefficient is about $\hat{\rho} = 64.5515$, thus $\hat{\rho}I = 64.5515*0.1248=8.0536$. On the other hand, the goodness of fit is $R^2=0.2301$, and the city number is $n=35$, thus $nR^2 = 35*0.2301 = 8.0536$. So, we have $\hat{\rho}I = nR^2$. This confirms equation (28) indirectly. The mean value of $n\mathbf{Wz}$ is $(\mathbf{Wz})^T\mathbf{o}=-0.1427$, which is just equal to the intercept of equation (5) (Table 1).

**Table 1 The parameter values and related statistics of spatial autocorrelation and spatial auto-regression models based on Moran's index**

| Measure | Year | Spatial autocorrelation model (equation 5) | | | Spatial autoregressive model (equation 8) | | | $R^2$ |
|---|---|---|---|---|---|---|---|---|
| | | Parameter | Coefficients | P-value | Parameter | Coefficients | P-value | |
| **City population** | 2000 | $(\mathbf{Wz})^T\mathbf{o}$ | -0.0839 | 0.0057 | $a$ | -0.1048 | 0.5861 | 0.0433 |
| | | $I$ | -0.0347 | 0.2303 | $\rho$ | -43.7192 | 0.2303 | |
| | 2010 | $(\mathbf{Wz})^T\mathbf{o}$ | -0.0916 | 0.0043 | $a$ | -0.0789 | 0.6878 | 0.0223 |
| | | $I$ | -0.0260 | 0.3914 | $\rho$ | -30.1294 | 0.3914 | |
| **Nighttime light** | 2000 | $(\mathbf{Wz})^T\mathbf{o}$ | -0.1255 | 0.0020 | $a$ | 0.1968 | 0.2944 | 0.1311 |
| | | $I$ | 0.0837 | 0.0325 | $\rho$ | 54.8709 | 0.0325 | |
| | 2010 | $(\mathbf{Wz})^T\mathbf{o}$ | -0.1427 | 0.0011 | $a$ | 0.2631 | 0.1404 | 0.2301 |
| | | $I$ | 0.1248 | 0.0035 | $\rho$ | 64.5515 | 0.0035 | |

**Note**: The values of spatial autocorrelation indexes and spatial autoregressive coefficients come between the intervals indicated by the maximum and minimum eigenvalues of spatial weight matrix and the derived matrixes.

The inherent relationships between spatial autocorrelation and spatial auto-regression are clearly reflected by the empirical analysis. If Moran's index is significant based on the confidence level of 95% ($P<0.05$), the corresponding spatial autoregressive coefficient is significant and *vice versa*. On the contrary, if Moran's index is not significant, the corresponding spatial autoregressive coefficient is not significant, too. Moran's index and the corresponding spatial autoregressive coefficient bear the same confidence level. For example, for NTL intensity (logarithmic variable) in 2010, both the $P$ value of Moran's index and that of the corresponding spatial autoregressive coefficient is $P=0.0035<0.05$. In contrast, for city population (logarithmic variable) in 2010, both the $P$ value of Moran's index and that of the corresponding spatial autoregressive coefficient is $P=0.3914>0.05$. This suggests that, for NTL intensity, spatial autocorrelation is significant, and corresponding spatial autoregressive modeling is meaning, while for city population, spatial autocorrelation is not significant, and corresponding spatial autoregressive modeling is meaningless. All these judgements are made on the basis of the significance level *α*=0.05.

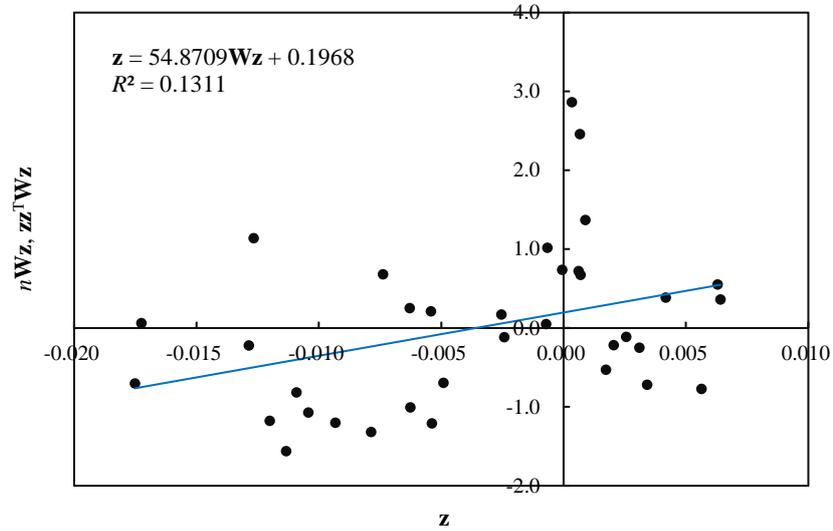

a. 2000

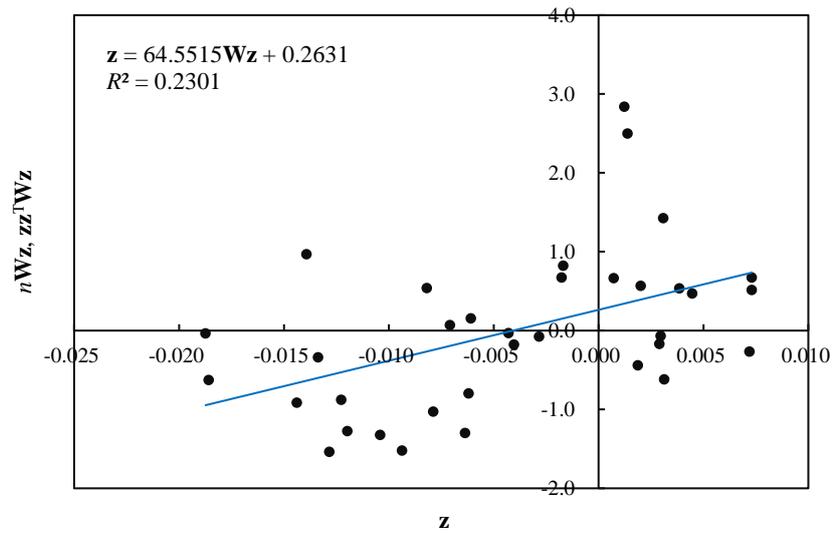

b. 2010

**Figure 1 Normalized scatterplots for spatial auto-regression of cities in Beijing-Tianjin-Hebei region, China**

Next, we investigate the significance of model parameters. Where slope parameter is concerned, a spatial autocorrelation model is significantly related to the corresponding spatial autoregressive model. However, as far as the intercept parameter, i.e., constant term, is concerned, it is hard to make it clear in a few words. For the spatial autocorrelation modeling results based on equation (6), all the intercept value are significant, that is, $P<0.05$. In contrast, for the spatial autoregressive modeling results based on equation (8), all the intercepts are not significant, i.e., $P>0.1$. The intercept of spatial autocorrelation model represent the mean value of $n\mathbf{Wz}$ or the sum of $\mathbf{Wz}$. The results indicate that the mean of $n\mathbf{Wz}$ cannot be ignored. If so, the mean of $\mathbf{Wz}$ cannot be ignored, too. In contrast, in light of equation (28) or (31), the intercept of the spatial autoregressive model is determined by three factors, namely, the mean of $n\mathbf{Wz}$, the value of Moran's index, and the goodness of fit. The goodness of fit is actually the squared coefficient of Pearson correlation between $\mathbf{z}$ and $\mathbf{Wz}$. If one of the above three factors is not significant, the intercept of the spatial autoregressive model will be not significant. This problem is important because it is involved with the derivation of the equation (4), which will be discussed in next section.

Finally, it is necessary to testify the key formulae derived above. Let's take the NTL intensity data of 2010 as an example. First, three quantities are prepared as follows. The inner product of $\mathbf{Wz}$ is $(\mathbf{Wz})^T(\mathbf{Wz})=0.0025$, the mean of $n\mathbf{Wz}$ is $(\mathbf{Wz})^T\mathbf{o}=-0.1427$, and the inner product of the city size and autoregressive model residuals is $\delta=\mathbf{z}^T\mathbf{\varepsilon}=26.9464$. According to equations (25) and (26), we have

$$\hat{a} = \frac{(n-\delta)(\mathbf{Wz})^T\mathbf{Wz} - I^2}{-I(\mathbf{Wz})^T\mathbf{o}} = \frac{(35-26.9464)*0.0025 - 0.1248^2}{-0.1248*(-0.1427)} = 0.2631, \quad (51)$$

$$\hat{\rho} = \frac{n-\delta}{I} = \frac{35-26.9464}{0.1248} = 64.5515. \quad (52)$$

According to equations (27), (28) and (31), we have

$$\hat{a} = -\frac{R^2}{I}(\mathbf{Wz})^T\mathbf{o} = -\frac{0.2301}{0.1248}*(-0.1427) = 0.2631, \quad (53)$$

$$\hat{\rho} = \frac{nR^2}{I} = \frac{35*0.2301}{0.1248} = 64.5515, \quad (54)$$

which also confirm equation (38). Finally, we can verify equations (33) and (34), the results are as below:

$$n(\mathbf{Wz})^T\mathbf{Wz} - \frac{I^2}{R^2} = 35*0.0025 - \frac{0.1248}{0.2301} = 0.0204, \quad (55)$$

$$((\mathbf{Wz})^T\mathbf{o})^2 = (-0.1427)^2 = 0.0204. \quad (56)$$

These calculation results confirm equation (33), and thus confirm equation (34) indirectly. In fact,

we can only directly testify the formula based on the computational world. As we know, the derived results based on the pure mathematical or theoretical world is not consistent with the observational data from the real world (Casti, 1996).

The above empirical analysis is made for testifying the theoretical derivation results rather than revealing the spatial pattern of urban system in the study area. In fact, the structure of spatial autoregressive model is not so satisfying. The soundness of model structure can be reflected by serial autocorrelation of residuals. Examining spatial error autocorrelation leads to one of ways of improving spatial autoregressive modeling (Haggett *et al*, 1977; LeSage, 2000; Permai *et al*, 2019; Ward and Gleditsch, 2008). Two measures can be employed to mirror serial autocorrelation of model residuals: one is Moran's index (Haggett *et al*, 1977), and the other is spatial Durbin-Watson statistic (Chen, 2016). The Moran's index of residuals sequence can calculated by the following formula

$$I_e = \mathbf{e}^\text{T}\mathbf{We}, \tag{57}$$

where **e** denotes the standardized residuals vector based on population standard deviation. The spatial Durbin-Watson statistic can be computed by the formula as below:

$$DW = 2C = \frac{2(n-1)}{n}(\mathbf{o}^\text{T}\mathbf{We}^2 - I_e), \tag{58}$$

in which *C* is Geary's coefficient of model's residuals sequence (Geary, 1954), *DW* denotes spatial Durbin-Watson statistic. The statistic information given by Moran's index is not consistent with that given by Durbin-Watson statistic (Table 2). There is evidence showing that Moran's index bears its limitations in estimating spatial dependence (Li *et al*, 2007). As far as city population is concerned, Moran's indexes of model residuals are not significant ($P>0.05$). Where NTL intensity is concerned, Moran's indexes of model residuals are significant to some extent ($P<0.05$, or $P\approx0.05$). However, all spatial Durbin-Watson statistics are not significant because all the results are close to the threshold value 2. According to the *DW* values, the spatial autoregressive models are acceptable based on the significance level of $\alpha=0.05$. However, according to Moran's index $I_e$ values, based on the significance level of $\alpha=0.05$, the spatial autoregressive model for 2000 is acceptable, but the model for 2010 is not acceptable. In short, the spatial autoregressive modeling effect is not satisfying. To research the urban system, we must improve the structure of spatial autoregressive model.

**Table 2 Moran's index and spatial Durbin-Watson statistics of the residual series of spatial**

**autoregressive models**

| Year | Statistics | City population | | NTL intensity | |
|---|---|---|---|---|---|
| | | Index value | $P$ value | Index value | $P$ value |
| **2000** | Moran's $I$ | 0.0464 | 0.2373 | -0.0404 | 0.0554 |
| | Spatial Durbin-Watson statistic | 1.7128 | | 1.9505 | |
| **2010** | Moran's $I$ | 0.0321 | 0.3950 | -0.0454 | 0.0199 |
| | Spatial Durbin-Watson statistic | 1.7687 | | 1.9881 | |

**Note**: For Durbin-Watson statistic based on $n=35$ and $\alpha=0.05$, the value of lower limit and upper limit are $d_l=1.402$ and $d_u=1.519$, respectively. So, $4-d_u=2.481$, $4-d_l=2.598$. If $DW<1.402$, the residuals bears positive sequential autocorrelation, and if $DW>2.598$, the residuals bears negative sequential autocorrelation. Now, the $DW$ values come between 1.519 and 2.481, indicating no significant sequential autocorrelation.

As indicated once and again, correlation analysis composes the basis of regressive analysis, and spatial auto-correlation and spatial auto-regression represent two different sides of the same coin. Although the aim of this study is not at the empirical research on spatial auto-regression of BTH cities, we might as well give the main conclusions drawn from data analysis. First, the spatial autocorrelation is nonlinear rather than linear relationships between these cities. For original variables, the spatial autoregressive coefficients are not significant; in contrast, if the size variables are taken logarithm, partial spatial autocorrelation coefficients are significant on the basis of probability $\alpha=0.05$. This indicates that the spatial autocorrelation and thus spatial auto-regression is based on allometric relationships instead of linear relationships. Second, for different measurements, spatial autoregressive effects are different. For population census data, spatial autoregressive coefficients are not significant; in contrast, for urban NTL data, spatial autoregressive coefficients are significant. This suggests that city population size and urban NTL intensity cannot be substituted for each other. Third, the intercept of spatial autoregressive models are not significant. This differs from the corresponding spatial autocorrelation models, in which the intercepts are significant for the probability $\alpha=0.05$. Fourth, the intercept of spatial autoregressive models influence the slope. This is also different from spatial autocorrelation model, in which the intercept does not influence slope value. Fifth, the evolution trend of spatial auto-regression of urban population sizes and NTL intensity is opposite. From 2000 to 2010, population spatial autoregressive parameters became more insignificant, while NTL spatial autoregressive parameters became more significant. This suggests that urban population variable is more suitable for conventional statistical analysis, while the NTL data is more suitable for spatial statistical analysis based on autocorrelation and auto-regression.

## 4. Discussion

By means of mathematical derivation and empirical analysis, a series of questions have been clarified. First, the inner product equation of Moran's index, equation (5), can be derived from a simple spatial autoregressive model, equation (7). The latter can be derived from the conventional spatial autoregressive model, which appeared in literature for a long time (Haggett *et al*, 1977; Odland, 1988; Whittle, 1954). The inner product equation is actually a spatial autocorrelation model, which is an inverse function of spatial autoregressive model and can be treated as an inverse spatial autoregressive model. Second, the linear regression analysis based on the least squares calculation is proved to be valid for estimating spatial autoregressive coefficients. Third, the value ranges of spatial autoregressive coefficient are derivable from two sides and three angles of view. Fourth, a group of useful relations can be derived from the spatial autoregressive models and autocorrelation formula. The relations have been utilized for deriving the value ranges of spatial autoregressive coefficient and will be useful in future studies on spatial autocorrelation and autoregressive analysis. The conversion relationship between spatial autocorrelation and spatial autoregressive modeling rests with reciprocal causation between a size variable and its spatial lag variable (Figure 2).

The novelty of this work lies in linking spatial autocorrelation measures and a spatial autoregressive model based on a standardized variable and normalized weight matrix in the simplest way. In theory, the product of Moran's index and spatial autoregressive coefficient is the number of elements, that is, $\rho I = \mathbf{z}^T \mathbf{z} = n$. In practice, the relation should be revised as $\rho I = nR^2$, where $R$ denotes the coefficient of Pearson correlation between size variable ($\mathbf{z}$) and spatial lag variable ($\mathbf{Wz}$). The mathematical derivation process for spatial autoregressive models can be generalized to Getis-Ord's index. Based on normalized variables and normalized weight matrixes, Getis-Ord's index can also be expressed as two equation based on inner product and outer product (Chen, 2020). Getis-Ord's index reflects spatial autocorrelation from another geographical prospective (Getis and Ord, 1992). Based on the inner product equation of Getis-Ord's index, another spatial autoregressive models can be constructed. By analogy, the inner product equation of Getis-Ord's index can be derived from a simple spatial autoregressive model based on normalized variable and globally normalized weight matrix.

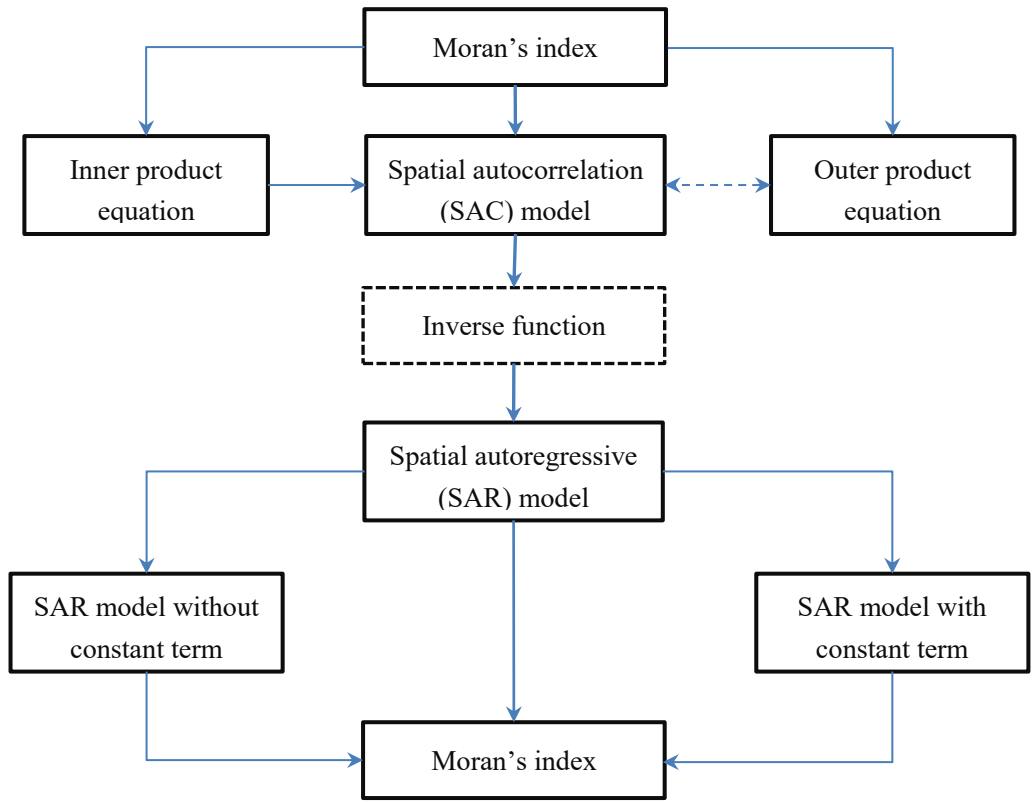

**Figure 2 Conversion relationship between spatial autocorrelation model and spatial autoregressive modeling**

**Note:** Moran's index can be decomposed into an inner product equation and an outer product equation. A spatial autocorrelation model can be constructed on the basis of the inner product equation. The outer product equation can be employed to improve Moran's scatter plot. The inverse function of the spatial autocorrelation model is actually a spatial autoregressive model, from which we can derive the inner product equation of spatial autocorrelation.

One of the aims of this work is at deriving the inner product equation of Moran's index. We can obtain the spatial autocorrelation model with constant term, that is, equation (4). However, we cannot deduce the model bearing no constant term, i.e., equation (5). Both equation (4) and equation (5) are based on the inner product of the size vector, **z**. If the constant terms in the simple spatial autoregressive models are removed, we can derive the inner product equation bearing constant term easily. In fact, the classical spatial autoregressive model bears no constant term (Odland, 1988). If the constant term in equation (7) is removed, we can derive equation (4) out of equation (7). The classical spatial autoregressive model came from the analogy with the temporal autoregressive models in the theory on time series analysis. In time series analysis, a temporal variable bears zero mean, or else it will be centralized so that its mean changes to zero. The formula of centralization is as follows: "*centralized variable = original variable – mean*". However, in spatial analysis, no evidence show that a spatial variable is a centralized variable. In this case, we should be cautious in dealing with the constant term in a spatial autoregressive model.

For the constant terms in the spatial autocorrelation model and spatial autoregressive models, we must pay attention to two facts. One fact is that the constant in the spatial autocorrelation model don't not influence the key parameter values. This has been proved in a companion paper. If the constant term in equation (5) is deleted, the slope representing Moran's index will not change. The trend line based on equation (5) is parallel to the trend line based on equation (3), while the trend line based on equation (4) coincides with equation (3). The other fact is that the constant term in the spatial autoregressive model is not often significant in the sense of statistics. This fact seems to lend further support to the form of classical spatial autoregressive model, that is, the constant term in the model is dispensable. As indicated above, the mean of $\mathbf{Wz}$ is $(\mathbf{Wz})^T\mathbf{o}/n$. Suppose that the independent variable $\mathbf{Wz}$ is centralized. From equation (18) it follows

$$\mathbf{z} = \frac{n}{I}(\mathbf{Wz} - \frac{1}{n}(\mathbf{Wz})^T\mathbf{o}), \tag{59}$$

which is the theoretical form of spatial autoregressive model based on centralized independent variable. Accordingly, from equation (27) it follows

$$\mathbf{z} = \frac{R^2 n}{I}(\mathbf{Wz} - \frac{1}{n}(\mathbf{Wz})^T\mathbf{o}) + \varepsilon, \tag{60}$$

which is the empirical form of spatial autoregressive model based on centralized independent variable. Obviously, the centralized variable $(\mathbf{Wz}-(\mathbf{Wz})^T\mathbf{o}/n)$ is significantly different from the non-centralized variable $(\mathbf{Wz})$.

The main shortcomings of this study lies in three aspects. First, as indicated above, how to understand and deal with the constant term in the spatial autoregressive model is a problem to be solved. Second, the inherent relationships between different sets of value ranges spatial autoregressive coefficient are not clear. As shown above, the first value range is based on the eigenvalues of the spatial weight matrix $\mathbf{W}$, the second value range is based on the eigenvalues of the inner product of the weight matrix, $\mathbf{W}^T\mathbf{W}$, and the third value range is based on the eigenvalues of the outer product of the weighted vector $\mathbf{Wz}$, that is, $\mathbf{W}^T\mathbf{zz}^T\mathbf{W}$. However, the relationships and differences between these ranges have not been deduced. Third, the mathematical reasoning is limited to Moran's index. Other spatial autocorrelation measurements and statistics such as Geary's coefficient and Getis-Ord's index has not been taken into account. As a matter of fact, there are rigorous relations between Moran's index and Geary's coefficient as well as Getis-Ord's index.

## 5. Conclusions

Spatial autocorrelation modeling is associated with spatial autoregressive modeling in geographical analysis. Using the simplest models based on a standardized variable and globally normalized weight matrix, we can reveal the mathematical and statistical relationships between spatial autocorrelation models and spatial autoregressive models. The key components of spatial autocorrelation and auto-regression can be abstracted as a pair of simple equations based on inner product and outer product. In this case, Moran's index and spatial autoregressive models can integrated into a canonical framework. In terms of the theoretical derivation and empirical analysis, the main points can be summarized as follows. *First, the inner product equation as a spatial autocorrelation model for estimating Moran's index can be derived from the simplest spatial autoregressive model and vice versa*. If there is spatial autocorrelation, we can construct a spatial autoregressive model based on standardized variables and normalized spatial weight matrixes. The normalized model is equivalent to the simplest classical spatial autoregressive model and can be verified by observational data. By mathematical transformation, we can derive the spatial correlation model from the spatial autoregressive model, and thus reveal the relationships between Moran's index and spatial autoregressive coefficient. *Second, the regression analysis based on the least square method is one of the effective ways to estimate spatial autoregressive coefficients*. The autoregressive coefficient can be derived from the least square formula. This suggests that spatial autoregressive coefficients can be estimated by the least square calculation. A number of empirical studies lend support to this judgment. This suggests that regressive analysis is indeed a valid approach to estimating spatial autoregressive parameters. *Third, there are more than one possible boundary values for spatial autoregressive coefficient*. By analogy with Pearson correlation coefficient, we know one set of the boundary values of Moran's index is -1 and 1. By analogy with Rayleigh quotient and related knowledge, we know another set of boundary values for Moran's index is *n* times of maximum and minimum eigenvalues. In terms of the relationships between Moran's index and spatial autoregressive coefficients, we can determine the boundary values for the coefficient of the basic spatial autoregressive model.

# Acknowledgements

This research was sponsored by the National Natural Science Foundation of China (Grant No. 42171192). The support is gratefully acknowledged.

# Appendixes

## Appendix 1. The parameter expressions of the linear model based on centralized variable and standardized variable

It can be proved that if the independent variable is centralized or standardized, the intercept of a linear model is equal to the mean of the dependent variable. As we know, a centralized variable equals the variable minus its mean. The centralized variable divided by its standard deviation is a standardized variable, i.e., $z$-score. Consider a simple linear regression model as below:

$$y = a + bx. \tag{A1}$$

The relation between the means of the two variables is as follows

$$\bar{y} = a + b\bar{x}. \tag{A2}$$

So, the intercept of equation (A1) is

$$a = \bar{y} - b\bar{x}. \tag{A3}$$

Subtracting $b\bar{x}$ from both sides of equation (A1) at the same time yields

$$y - b\bar{x} = a + bx - b\bar{x}. \tag{A4}$$

Substituting equation (A2) into equation (A4) yields

$$y = a + b\bar{x} + b(x - \bar{x}) = \bar{y} + b(x - \bar{x}) = \bar{y} + (s_x b)\frac{x - \bar{x}}{s_x}, \tag{A5}$$

where $s_x$ denotes the standard deviation of $x$. This indicates that the intercept of the linear model is just the mean of the dependent variable $y$.

## Appendix 2. The relation between the slope of a linear model and the slope of its inverse function

Theoretically, the slope of a univariate linear regression model is the reciprocal of the slope of its inverse function. However, in empirical analysis, the product of the slope of a univariate linear regression model and the slope of its inverse function equals the determination coefficient ($R^2$). Based on the mathematical world, a univariate linear regression model can be expressed as

$$y = a + bx, \tag{B1}$$

where $x$ denotes independent variable, $y$ refers to dependent variable. As for the parameters, $a$ is the intercept, and $b$ is the slope. The inverse function is

$$x = a' + b'y = -\frac{a}{b} + \frac{1}{b}y, \tag{B2}$$

where $a'=-a/b$, $b'=1/b$, and thus $bb'=1$. However, based on the real world and computational world, the regressive model based on $n$ sample points should be expressed as

$$y_i = a + bx_i + \varepsilon_i, \tag{B3}$$

where $\varepsilon$ represents residuals ($i=1,2,3,\ldots,n$). The inverse function is

$$x_i = a' + b'y_i + \varepsilon'_i, \tag{B4}$$

in which $a'$, $b'$ refers to parameters, and $\varepsilon'$ denotes residuals. In this case, $bb'\neq 1$. By the idea from the least square sum of errors, the slope of equation (B3) can be derived as

$$b = \frac{\sum_{i=1}^{n}(x_i - \bar{x})(y_i - \bar{y})}{\sum_{i=1}^{n}(x_i - \bar{x})^2} = \frac{\text{cov}(x,y)}{\text{Var}(x)} = \frac{\text{cov}(x,y)}{\sigma_x \sigma_y}\frac{\sigma_y}{\sigma_x} = R\frac{\sigma_y}{\sigma_x}, \tag{B5}$$

in which Var ($\cdot$) is a variance function, cov($\cdot,\cdot$) is a covariance function, $\sigma$ is standard deviation, and $R$ is multiple correlation coefficient. Accordingly, the slope of equation (B3) can be derived as

$$b' = \frac{\sum_{i=1}^{n}(x_i - \bar{x})(y_i - \bar{y})}{\sum_{i=1}^{n}(y_i - \bar{y})^2} = \frac{\text{cov}(x,y)}{\text{Var}(y)} = \frac{\text{cov}(x,y)}{\sigma_x \sigma_y}\frac{\sigma_x}{\sigma_y} = R\frac{\sigma_x}{\sigma_y}. \tag{B6}$$

Multiply equation (B5) and equation (B6) yields

$$bb' = R^2 \frac{\sigma_y}{\sigma_x}\frac{\sigma_x}{\sigma_y} = R^2. \tag{B7}$$

So we have

$$b' = \frac{R^2}{b}. \tag{B8}$$

The above process of mathematical reasoning is suitable for the spatial autoregressive models,

equations (7) and (8).